\documentclass[11pt,onecolumn]{article} 
\usepackage{amsmath}
\usepackage{amsfonts}
\usepackage{graphicx}
\usepackage[latin1]{inputenc}
\usepackage{cite}
\usepackage{color}
\usepackage{hyperref}

\begin{document}

\baselineskip24pt

\thispagestyle{empty}

\vspace*{2cm}

{\noindent \Large \bf  Cochlear-bone wave can yield a hearing sensation as well as otoacoustic emission}\\

{\noindent Tatjana Tchumatchenko$^{1}$ and Tobias Reichenbach$^{2,\ast}$}\\

{\noindent $^1$ Department Theory of Neural Networks, Max Planck Institute for Brain Research, Max-von-Laue Str.~4, 60438 Frankfurt am Main, Germany \\
$^2$ Department of Bioengineering, Imperial College London, South Kensington Campus, London SW7 2AZ, U.~K.}\\

{\noindent $^\ast$ To whom correspondence should be addressed; E-mail:  reichenbach@imperial.ac.uk.}

\newpage

{\bf 

A hearing sensation arises when the elastic basilar membrane inside the cochlea vibrates. The basilar membrane is typically set into motion through airborne sound that displaces the middle ear and induces a pressure difference across the membrane. A second, alternative pathway exists, however: stimulation of the cochlear bone vibrates the basilar membrane as well. This pathway, referred to as bone conduction, is increasingly used in the construction of headphones that bypass the ear canal and the middle ear. Furthermore, otoacoustic emissions, sounds generated inside the ear and measured in the ear canal, may not involve the usual wave on the basilar membrane, suggesting that additional cochlear structures are involved in their propagation. Here we describe a novel propagation mode that emerges through deformation of the cochlear bone. Through a mathematical and computational approach we demonstrate that this wave can explain bone conduction as well as numerous properties of otoacoustic emissions.
}

\newpage

\section*{Introduction}

The mammalian cochlea is an intricate device that acts as a spatial frequency separator~\cite{Pickles,lighthill-1981-106,Ulfendahl1997331,robles-2001-81}. Airborne sound vibrates the middle ear and evokes a pressure signal at the base of the fluid-filled inner ear (Figure~\ref{MiddleEar}). The pressure oscillation then propagates as a surface wave on the basilar membrane, an elastic structure that separates two fluid-filled compartments in the cochlea. Different frequency components become spatially separated  because, through changes in its material properties, the basilar membrane is tuned to a range of  frequencies that systematically vary between the apical and the basal end. A segment of the basilar membrane near the base resonates at a high frequency, and segments from further apical positions resonate at successively lower frequencies. The wave on the basilar membrane elicited by a single frequency greatly  increases in amplitude upon approaching its resonant position, beyond which it sharply declines~\cite{lighthill-1981-106,robles-2001-81}.  A tonotopic map emerges in which high frequencies are detected near the base and low frequencies near the apex of the cochlea.

The basilar-membrane waves produced by different frequencies, however, do not simply superpose linearly. Instead, the basilar membrane at a given cochlear position responds nonlinearly to forcing near the resonant frequency of that location~\cite{robles-2001-81,Ulfendahl1997331}. The nonlinearity arises from mechanical activity of hair cells that reside on the basilar membrane. These cells can produce mechanical forces that greatly amplify weak stimuli; large vibrations are amplified less. The relation between the amplitude of the applied force and the resulting vibration is hence compressively nonlinear, and indicates that each basilar-membrane segment operates near a dynamic instability (Hopf bifurcation)~\cite{eguiluz-2000-84,camalet-2000-97,hudspeth-2010-104}. 

The nonlinear response of the basilar membrane produces distortion when multiple pure tones are presented simultaneously~\cite{ruggero-1991-349,robles-1997-77,cooper-1998-509,julicher-2001-98}. As an example, a cubic nonlinearity yields a response at frequencies such as $2f_1-f_2$ or $2f_2-f_1$ when  stimulated at two frequencies $f_1$ and $f_2$. Such distortion products indeed arise prominently in the cochlea. Because they can not only be measured as basilar-membrane vibration, but also with a microphone placed in the ear canal, they must be emitted from the cochlea into the ear canal. One accordingly refers to these tones as distortion-product otoacoustic emissions.

For a given frequency, the peak of the traveling wave is relatively sharp, with a longitudinal extent of only around $0.5$~mm~\cite{Ulfendahl1997331,robles-2001-81}.   The cubic distortion frequencies $2f_1-f_2$ or $2f_2-f_1$, for instance, are therefore only created at a significant amplitude when the two primaries $f_1$ and $f_2$ are sufficiently close, such that the corresponding peak regions overlap. The distortion hence arises from a narrow cochlear region from which it must propagate back to the base to cause a sound signal in the ear canal.

How the backward propagation occurs is currently intensely debated. Experiments show that distortion-product otoacoustic emissions consist of two components that differ in the temporal delay between their generation and the resulting emission in the ear canal~\cite{knight-2000-107,knight-2001-109}. One component has a long delay of a few milliseconds, whereas the delay of the other component is much shorter. The delay is measured through the change in phase of an emission upon altering the primary frequencies.

Some theoretical studies have suggested that both components emerge through waves on the basilar membrane that propagate backward from their generation site to the cochlear base~\cite{zweig-1995-98,shera-1999-105,kalluri-2001-109}. Measurements of the intracochlear pressures as well as the cochlear microphonic potential support such reverse basilar-membrane waves~\cite{dong-2008-123,meenderink-2010-103}.  Recent experimental measurements that have directly recorded the waves propagating along the membrane, however, only found forward-traveling waves, both at the primary frequencies as well as at the distortions~\cite{Ren2004,hea-2007-228,He2008}. Moreover, the stapes appear to vibrate at the distortion signal  before the basilar membrane.

Recently we have proposed that the long-delay component of a distortion-product otoacoustic emission arises through waves on Reissner's membrane, another elastic membrane within the cochlea that extends in parallel to the basilar membrane from the cochlear base to the apex~\cite{reichenbach-2012-1}. Our theoretical and numerical considerations show that short surface waves  can propagate along Reissner's membrane, and that those waves can  be created through the cochlear active process. Laser-interferometric measurements performed by ourselves have confirmed that such waves on Reissner's membrane exist and can arise from distortion on the basilar membrane.

Because waves on Reissner's membrane have relatively short wavelengths, below $0.5$~mm for frequencies above a few kHz, such backward-propagating waves have slow speeds of a few meter per second. Distortion products emerging through those waves yield accordingly  delays of a few milliseconds when propagating from their generation region to the middle ear.

How the short-delay component of an otoacoustic emission emerges, if not through backward waves on the basilar membrane, remains elusive. It has been suggested that compressional waves may transport a distortion signal within the cochlea~\cite{Ren2004,hea-2007-228,He2008}. Indeed, such waves can propagate in the cochlear fluids at large wavelengths and speeds. Because they involve no pressure difference across the basilar membrane and hence no membrane vibration, however, they cannot be produced by hair-cell forces acting on the membrane. Instead, their generation would require the active process to produces local volume changes, which have not yet been detected.

The mechanism of signal transmission in bone conduction remain similarly elusive. Bone conduction refers to our ability of hearing auditory signals through vibration of the cochlear bone, even in the absence of a functional middle ear~\cite{tonndorf-1976-5}. Already one of the pioneers of hearing research, G.~v.~B\'{e}k\'{e}sy, conducted experiments in which he showed that the hearing sensation that is produced through bone conduction can be canceled by stimulating the ear by an identical, but airborne, signal when its amplitude and phase are chosen carefully~\cite{Bekesy}. Bone conduction hence appears to elicit the same basilar-membrane wave as is produced by airborne sound. This way of stimulating the ear is now increasingly used for constructing speakers, for example bone-conduction headphones such as in the novel  Google Glass device, that vibrate the cochlear bone and do not obstruct the ear canal. Such headphones allow to listen to environmental sound and, for example, additional information such as navigational directions that are inaudible to others. Despite the increasing use of this technology, we lack an understanding of how the cochlear bone vibration leads to basilar-membrane waves and hence the hearing sensation.

Early studies by B\'{e}k\'{e}sy as well as Herzog and Krainz suggested that the cochlear bone may not just vibrate homogeneously but deform under sound vibration~\cite{Bekesy,Herzog1926}. If the basilar membrane was not positioned in the middle of the cochlea, bone deformation could deflect the membrane and hence elicit the well-known basilar-membrane wave.

In this article we employ a  cochlear model to show that deformation of the bone produces a wave that travels along the bone and that couples to the basilar membrane. 
Through mathematical and numerical methods we investigate whether this wave can underlie bone conduction as well as transmit distortion products created within the cochlea back to the ear canal. 

We find that the deformation of the cochlear bone that can be triggered by direct bone stimulation evokes a traveling wave long the basilar membrane. We also show that otoacoustic emissions emerging from the cochlea via bone conducted waves can have short delays on the order of a few milliseconds. These results shed light on the mechanisms behind bone conduction and otoacoustic emissions and thereby can help advance their commercial and clinical applications.

\newpage

\section*{Results}

\begin{figure}[h]
\includegraphics[width=\columnwidth]{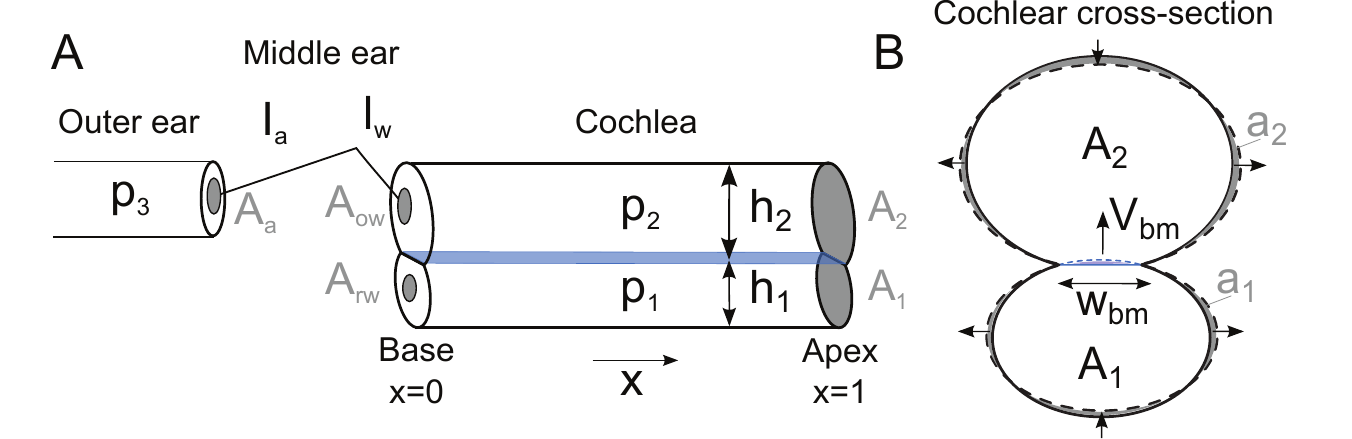}
\caption{{\bf{Anatomy of the outer and inner ear}} {(\bf A}) Sound causes a pressure vibration $p_3$ in the ear canal and a motion of the ear drum (area $A_\text{a}$). The middle ear's ossicles, namely the mallus of  length $l_\text{a}$, incus of length $l_\text{w}$, and stapes convey the motion to the inner ear, or cochlea, to vibrate the oval window (area $A_\text{ow}$) and the round window (area $A_\text{rw}$). The pressures $p_1$ in the scala tympani and $p_2$ in the  scala vestibuli change accordingly. ({\bf B}) A transverse section of the inner ear shows the basilar membrane separating two chambers of cross-sectional area $A_1$ and $A_2$. Vibration of the membrane (velocity $V_\text{bm}$) and deformation of the cochlear bone, at constant circumference, lead to area changes $a_1$ and $a_2$.
\label{MiddleEar}}
\end{figure}

We start from a one-dimensional model of the inner ear (Figure~\ref{MiddleEar}). The basilar membrane extends in the longitudinal $x$-direction and delineates two chambers. The one below the membrane is the scala tympani. We denote a pressure deviation therein from the resting pressure by $p_1$, a longitudinal fluid flow by $j_1$, and the cross-sectional area by $A_1$.  The upper chamber comprises the scala media and scala vestibuli; this chamber's pressure deviation is $p_2$, its longitudinal fluid flow $j_2$, and its cross-sectional area $A_2$.

The longitudinal fluid flow in the upper and lower chamber carry a momentum $\rho\partial_t j_1$ and  $\rho\partial_t j_2$, respectively, which must result from a longitudinal pressure gradient in that chamber:
\begin{align}
\rho\partial_t j_1&=-A_1\partial_x p_1,\cr
\rho\partial_t j_2&=-A_2\partial_x p_2.
\label{eq:momentum}
\end{align}
Here $\rho$ denotes the fluid density. The continuity equation states that 
a gradient in the longitudinal fluid flow of either chamber can only arise from a temporal change in the chamber's cross-sectional area or from a change in the fluid's density $\rho_{1/2}$. Denote by $a_1$ and $a_2$ the area change of the upper respectively  the lower chamber, such that the total cross-sectional area of the respective chamber is $A_{1/2}+a_{1/2}$. We then find
\begin{align}
\partial_x j_1+\partial_t a_1+\frac{A_1}{\rho_1}\partial_t\rho_1=0,\cr
\partial_x j_2+\partial_t a_2+\frac{A_2}{\rho_2}\partial_t\rho_2=0.
\label{eq:continuity}
\end{align}
A deviation in the fluid's density  from its resting value $\rho_0$ is caused by a change in pressure through the fluid's compressibility $\kappa$: $\partial_t\rho_{1/2}=\rho_0\kappa\partial_tp_{1/2}$. 

The cross-sectional area of either cochlear chamber can change because of  basilar-membrane vibration (Figure~\ref{MiddleEar}B). We assume that the membrane's cross section deforms parabolically, with a midpoint velocity $V_\text{bm}$ that is defined such that an upward membrane motion yields a positive velocity (Methods). This motion  hence expands the lower chamber and shrinks the upper one:
\begin{equation}
\partial_t a_1=-\partial_t a_2=\frac{2}{3}\cdot w_\text{bm}\cdot V_\text{bm},
\label{eq:area_change}
\end{equation}
in which $w_\text{bm}$ denotes the membrane's width. In the following we consider a sound signal at a single  angular frequency $\omega$. Pressure vibration occurs at that same frequency, and we make an ansatz in which it propagates longitudinally with a wave vector $k$ and an amplitude $\tilde{p}_{1/2}$:
\begin{equation}
p_{1/2}=\tilde{p}_{1/2}e^{i\omega t-i k x} + \text{c.c.}.
\label{Ansatz}
\end{equation}
Hereby c.c. denotes the complex conjugate. Similarly, the basilar-membrane velocity oscillates at frequency $\omega$ and propagates longitudinally, it can hence be written as
\begin{equation}
V_\text{bm}=\tilde{V}_\text{bm}e^{i\omega t-i k x} + \text{c.c.}.
\label{Ansatz_V}
\end{equation}
We can now relate the difference of the pressure amplitudes across the basilar membrane to the vibrational amplitude that it evokes:
\begin{equation}
\tilde{p}_1-\tilde{p}_2=Z_\text{bm}\cdot \tilde{V}_\text{bm}.
\label{MembraneBoundary1}
\end{equation}
The coefficient $Z_\text{bm}$ denotes the local acoustic impedance of the membrane, which in general depends on the frequency of stimulation, see Tab.~\ref{Parameters}. The equation~(\ref{eq:momentum}) of momentum together with the equation~(\ref{eq:continuity}) of continuity  and the equations~(\ref{eq:area_change}) and (\ref{MembraneBoundary1}) for the basilar-membrane velocity yield the well-known cochlear waves  that propagate along the basilar membrane.

Here, we also include the possibility that the cochlear bone around the upper and lower chamber can be deformed through the intra-chamber pressure. Two types of deformation of a given cross-section  are conceivable. First, the circumference of a cross-section may change. This requires compressibility of the chamber's wall. Second, the circumference may remain constant but  the shape of the cross-section may vary. Because the elastic modulus of bone is high, the first type of deformation has a much higher impedance than the second~\cite{{Reilly,TimoshenkoGere}}. We hence only consider a deformation of the second type.

Which change in cross-sectional area results from a deformation that leaves the circumference constant? Let us approximate each chamber's cross-section by an ellipse-like shape that is deformed under internal pressure (Figure~1B and section 4.2 in methods). Because a pressure change produces an equal force at every angle, no deformation can result when the cross section is circular (and when it where to remain circular with an identical circumference). An asymmetric ellipse-like object that lacks rotational symmetry, however, will deform under a pressure changes. The impedance associated with this deformation has been studied in the literature~\cite{TimoshenkoGere}. Specifically, a decreasing internal pressure will increase the asphericity of the ellipse-like shape because, at constant circumference, the  area is the smaller the more aspherical it is. Conversely, an enhanced pressure will tend to increase the cross-sectional area, which will hence deform towards a circle. For small deformations as we consider here, the area change depends linearly on the pressure deviation. The total change $a_{1/2}$ of the cross-sectional area of the upper respectively the lower chamber is hence the sum of one contribution from the membrane deflection and another contribution from the bone deformation:
\begin{align}
\widetilde{a}_1&=-\frac{2i}{3\omega}\cdot w_\text{bm}\cdot \tilde{V}_\text{bm}+C\cdot \tilde{p}_1,\cr 
\widetilde{a}_2&=\frac{2i}{3\omega}\cdot w_\text{bm}\cdot \tilde{V}_\text{bm}+C\cdot \tilde{p}_2.
\label{eq:area_deformation}
\end{align}
 $C$ is a linear-response coefficient that we assume to be identical for both chambers. Its value can be derived through computing the elastic deformation of a tube (Methods section 4.2 as well as Ref.~\cite{TimoshenkoGere}, pp. 289-296) and is given by 
\begin{equation}
C=\frac{4\pi(1-\nu^2)}{E}\frac{R^3 w_0^2}{h^3}.
\label{eq:linear_response}
\end{equation}
Here, $E$ denotes the Young's modulus of the cochlear bone, $\nu$ the Poisson ratio, $h$ the thickness of the cochlear bone, $R$ the average  radius of a chamber and $w_0$ the (approximately elliptical) deformation of the cross-sectional shape, see Tab.~\ref{Parameters}.

The fluid-momentum equation~(\ref{eq:momentum}) with the continuity equation~(\ref{eq:continuity}) as well as equations~(\ref{MembraneBoundary1}) and (\ref{eq:area_deformation}) for area and membrane vibration yield the matrix equation
\begin{equation}
 k^2 \left(\begin{array}{c} \tilde{p}_1 \\ \tilde{p}_2\end{array} \right) = \mathcal{M} \left(\begin{array}{c} \tilde{p}_1 \\ \tilde{p}_2\end{array} \right) 
 \label{eq:matrix_wave}
 \end{equation}
 with the $2\times 2$ matrix 
\begin{equation}
\mathcal{M} =-\omega \rho_0\left(\begin{array}{cc}  \frac{2i w_\text{bm}}{3A_1Z_\text{bm}}-\frac{\omega C}{A_1} -\omega\kappa& -\frac{2i w_\text{bm}}{3A_1 Z_\text{bm}}\\[6pt]
-\frac{2i w_\text{bm}}{3A_2 Z_\text{bm}} &  \frac{2i w_\text{bm}}{3A_2Z_\text{bm}}-\frac{\omega C}{A_2}-\omega\kappa
 \end{array} \right).
 \label{Dispersion}
\end{equation}
The possible wave vectors $k$ hence follow from the eigenvalues of the matrix $\mathcal{M}$. The eigenvectors describe how the pressures in the upper and lower chamber relate to each other in the corresponding wave mode.

The eigenvalues and eigenvectors can be readily interpreted for the different terms in the matrix $\mathcal{M}$ are of different orders of magnitude:  $|w_\text{bm}/(A_{1/2}Z_\text{bm})|\gg |\omega C/A_{1/2}|\gg\omega\kappa$.  The basilar membrane is significantly floppier than the cochlear bone, and yields a dominating contribution in the matrix $\mathcal{M}$. The effect of the fluid's compressibility is negligible. In the following we hence regard the fluid  as incompressible.

Because the matrix equation~(\ref{eq:matrix_wave}) has two degrees of freedom, there exist two eigenvectors that correspond to two distinct wave modes.
First, one eigenvector involves opposite pressures in the two chambers,  $A_2\tilde{p}_{1}=-A_1\tilde{p}_2$, and yields a wave vector  
\begin{equation}
k_\text{bm}=\pm\sqrt{-\frac{2i\rho \omega w_\text{bm}}{3 Z_\text{bm}} \left(\frac{1}{A_1}+\frac{1}{A_2}\right)}.
\label{eq:kbm}
\end{equation}
This wave vector does not involve deformation of the cochlear bone. Instead, it  follows from the basilar-membrane impedance $Z_\text{bm}$ alone and yields the well-known basilar-membrane wave. 

Because the basilar-membrane impedance varies longitudinally, the wave's amplitude changes as well. A Wentzel-Kramers-Brillouin (WKB) approximation can be applied and  reveals that the local wave vector still follows from Equation~(\ref{eq:kbm}), whereas the  pressure amplitude is proportional to the inverse square root of the wave vector (Methods).

Second, and most important for our study here, the other eigenvector,  $\tilde{p}_{2}=\tilde{p}_{1}$, involves pressures in both chambers that are equal at any given longitudinal location. The corresponding wave accordingly does not deflect the basilar membrane. It solely evokes deformation of the cochlear bone that propagates at a wave vector $k_\text{cb}$:
\begin{equation}
k_\text{cb}=\pm\sqrt{\frac{2\rho \omega^2 C}{A_1+A_2}}.
\end{equation}
We refer to this mode as the cochlear-bone wave. Because the impedance of the cochlear bone remains approximately constant between the cochlear base and apex, this wave's amplitude remains constant as well and we do not need to employ the WKB approximation. The wavelength is the longer the larger the impedance of the cochlear bone. Because this impedance is relatively high it yields a comparatively long wavelength, on the order of a few millimeter to a few centimeter, and accordingly a propagation speed that exceeds that of the basilar-membrane wave. Notably, the wavelength of the cochlear-bone wave is still substantially below that of a compressive fluid wave which reflects the above finding that the fluid compressibility plays a negligible role.  

Although both basilar and cochlear waves are clearly distinct, with one wave depending only on the basilar-membrane impedance and the other wave solely on the impedance of the cochlear bone, they couple in two intriguing ways. One type of coupling becomes important for otoacoustic emissions and the other for bone conduction.

As the first type of coupling, a force that acts on the basilar membrane can elicit the cochlear-bone wave. This unexpected effect becomes clear when we recall that a displacement of the basilar membrane increases the pressure in one chamber but decreases it in the other by the same amount, $\tilde{p}_1=-\tilde{p}_2$. Such displacement can elicit a wave on the basilar membrane that  involves opposite pressures, $A_2\tilde{p}_1=-A_1\tilde{p}_2$. In an asymmetric cochlea, with $A_1\neq A_2$, the pressure changes evoked by basilar-membrane motion do not fully match those involved in the basilar-membrane wave.  A force that acts on the basilar membrane must hence, besides the basilar-membrane wave, stimulate a second degree of freedom: the wave on the cochlear bone. Because otoacoustic emissions arise from the activity of hair cells on the basilar membrane, they can hence excite a cochlear-bone wave and thus propagate out of the cochlea. Below we show this mechanism in detail for the case of distortion-product otoacoustic emissions.

In the second, in a sense reverse way of coupling, stimulation of the cochlear bone can elicit a basilar-membrane wave. Assume that, at a certain longitudinal location,  both cochlear chambers change their area by the same amount due to forcing. Because the cross-sectional areas of both chambers are, in general, different, such forcing produces different pressures in the two chambers and hence a displacement of the basilar membrane. This mechanism can yield bone conduction as we show below.

\subsection*{Distortion products}

Distortion products are combination tones that the cochlea produces when it encounters multiple frequencies. 
As a prominent example, when stimulated by two close frequencies $f_1$ and $f_2$, in which $f_1$ is smaller than $f_2$ by convention, the inner ear yields emissions at cubic distortion frequencies such as $2f_1-f_2$ and $2f_2-f_1$.

This distortion is produced by a  nonlinearity on the basilar membrane. Indeed, close to its resonant position, the linear response~(\ref{MembraneBoundary1}) of the basilar membrane is supplemented by a cubic nonlinearity that originates in the amplification provided by hair cells:
\begin{equation}
\tilde{p}_1-\tilde{p}_2=Z_\text{bm}\cdot \tilde{V}_\text{bm} +3A\widetilde{V_\text{bm}^3}
\label{MembraneBoundary_nonlinear}
\end{equation}
in which $A$ is a coefficient. Distortion arises for the Fourier transform of the cubic nonlinearity can be written as the convolution of Fourier coefficients: $\widetilde{V_\text{bm}^3}=\tilde{V}_\text{bm} * \tilde{V}_\text{bm} * \tilde{V}_\text{bm}$, which yields mixing in the frequency domain.

To solve the  nonlinear equation~(\ref{MembraneBoundary_nonlinear}), we first compute Green's functions, that is pressures $\tilde{p}_{1/2}^G(x,x_0,\omega)$ that result from a single force at position $x_0$:
\begin{equation}
\tilde{p}_1^G(x,x_0,\omega)-\tilde{p}_2^G(x,x_0,\omega)=Z_\text{bm}\cdot \tilde{V}_\text{bm} +p_F\cos(\omega t)\delta(x-x_0).
\label{eq:green}
\end{equation}
Using techniques from complex analysis, we obtain an analytical solution for these Green's functions (Methods). The solution consist of two waves modes, the basilar-membrane wave as well as the wave on the cochlear bone. The latter is excited when the cochlear chambers are asymmetric, $A_1\neq A_2$. In this case, the nonlinear basilar-membrane response accordingly  produces not only a basilar-membrane wave, but also a cochlear-bone wave.

Within each wave mode, two distinct waves emerge. First, one wave travels backward from the generation site $x_0$ to the stapes. The  second wave moves forward to the apex. Although it may undergo reflection at the apex, we ignore this forward-traveling wave in the following and only consider the wave that travels backward.

Because the cochlear nonlinearity extends over a certain region near the peaks of the primary frequencies, many  such waves are produced and add up to yield the net distortion product. Mathematically this follows from  integrating the Green's functions~(\ref{eq:green}) together with the nonlinear inhomogeneity $\widetilde{V_\text{bm}^3}$, which yields the solution to the inhomogeneous differential equation~(\ref{MembraneBoundary_nonlinear}):
\begin{equation}
\tilde{p}_{1/2}(x,\omega)=\frac{3A}{p_F}\int_0^Ldx_0  \tilde{p}_{1/2}^G(x,x_0,\omega)\widetilde{V_\text{bm}^3}(x_0,\omega).\label{eq:greenConv}
\end{equation}

What happens to the backward-traveling waves, the one in the basilar-membrane and the other in the cochlear-bone mode? Part of the energy that they carry will be emitted into the ear canal. The remainder will be reflected off the middle ear and produce forward traveling waves. One such wave will propagate on the basilar membrane, and the other as cochlear-bone deformation.

The reflection of the backward-propagating waves off the middle ear can be quantified by considering the action of the middle ear (Methods). Indeed, the middle ear acts as an impedance transformer to match the impedance of an incoming sound to that of the basilar-membrane wave. An incoming sound is hence largely transmitted to basilar-membrane motion, without much reflection at the middle ear. Reversely, a backward-propagating basilar-membrane wave is effectively transmitted to a sound wave, and not much reflection occurs. A backward-propagating cochlear-bone wave, in contrast, will be much less transmitted for its impedance differs from the basilar-membrane wave and is not matched by the middle ear. Considerable reflection then occurs and produces forward-traveling waves,  in particular a wave on the basilar membrane.

Three basilar-membrane waves hence propagate  at the distortion frequency (Figure~\ref{ForwardBackward}A). First, a forward-traveling wave is generated by the basilar-membrane's nonlinearity. This wave is predominantly created in the region where the primary frequencies overlap. Because the contributions from this region differ in phase, they partly cancel, and the wave has an amplitude peak at the point of maximal generation.  For the lower sideband distortion frequency $2f_1-f_2$ that we consider here, the wave then travels further apical and experiences a second peak near its resonant position.

Second, the nonlinear basilar-membrane response creates a backward-propagating wave as well. As for the forward-traveling wave, the contributions to this wave from different cochlear locations partly annihilate each other, and the amplitude of this wave is largest at the point of maximal generation. The wave cannot be created apical to the resonant position of the upper primary frequency, $f_2$, such that no backward wave arises there.

Third, a reflected forward-traveling wave arises from the reflection of the reverse basilar-membrane and the cochlear-bone wave. This wave's amplitude behaves as the usual basilar-membrane wave: its amplitude increases until it reaches its resonant position, beyond which it sharply diminishes.

The first and third component superimpose to yield the net forward-traveling wave on the basilar membrane. Can that wave have a larger amplitude than the reverse basilar-membrane wave and hence conceal its existence? 

Our numerical simulations show that the answer depends on the ratio of the primary frequencies as well as, potentially, on the cochlear location (Figure~\ref{ForwardBackward}B). When the primary frequencies are sufficiently apart, the reverse wave can blanket the forward-propagating waves. Close primary frequencies, however, yield a net forward-traveling wave that exceeds the backward-propagating one at all cochlear locations.

In order to intuitively understand these results, we recall that the distortion  is generated within an extended cochlear region, namely where the peaks of the primary-frequency waves  significantly overlap. The phase of the  distortion changes with location, and the produced reverse-propagating waves hence experience significant destructive interference. This destructive interference is the stronger the faster the phases change, and hence the smaller the wavelength is. Generation close to the peak region, where the basilar-membrane wave is short,  yields accordingly more destructive interference then generation more apical. Similarly, because the cochlear-bone wave has a comparably long wavelength, its generation comes with less destructive interference than  that of the basilar-membrane wave.

Because the basilar-membrane waves of closer primary frequencies overlap stronger, they produce more destructive interference in the generated, reverse basilar-membrane wave. In relation to the latter the produced  backward-traveling cochlear-bone wave is therefore stronger and yields accordingly a stronger reflection. Part of that reflection is a forward-traveling basilar-membrane wave which hence blankets the reverse wave on the basilar membrane.

\subsection*{Bone conduction}

Deformation of the cochlear bone can elicit basilar-membrane waves and hence a hearing sensation. Similarly to our calculations regarding distortion-product otoacoustic emissions, we quantify this effect through computing Green's functions, that is the pressure waves that result from deforming the cochlear bone at a single longitudinal location $x_0$ (Methods). Specifically, we consider a deformation of the cochlear bone such that the cross-sectional area of the upper chamber vibrates in phase with that of the lower chamber, and with the same amplitude.

The Green's functions show that four waves emerge from such stimulation: two cochlear-bone waves, traveling basally and apically from the stimulation site, and two basilar-membrane waves, also propagating backward and forward. The basilar-membrane waves are hereby only excited if the two chambers differ in their cross-sectional area, $A_1\neq A_2$. In a hypothetical symmetric inner ear, in which the areas are equal, deformation of the cochlear bone would not elicit basilar-membrane waves, as had already been remarked by B\'ek\'esy~\cite{Bekesy}.

We are interested in the basilar-membrane waves because they elicit the hearing sensation. Apical to the stimulation point, we find a forward traveling wave that peaks close to its resonant position and resembles the standard, middle-ear-evoked waves for all stimulation points (Figure~\ref{WallStim}A). Basal to the stimulation point we obtain a backward-traveling wave that decays in amplitude as it travels towards the base. The amplitude of the elicited basilar-membrane wave depends on the stimulation position along the cochlea: it increases for more basal stimulation. The shape of the produced wave is, however, largely independent of the location of stimulation. Compressive stimulation of an extended region of the cochlear bone generates a superposition of the waves elicited by point stimulation. The extent of the stimulation region governs the amplitude but not the spatial profile of the basilar-membrane motion.
 
The amplitude of the elicited basilar membrane motion depends on the impedance of the cochlear bone as compared to the membrane's (Fig.~\ref{WallStim}B). The impedance associated to bone deformation is generally higher than that of the basilar membrane. The smaller the bone's impedance, the more similar it becomes to that of the membrane.  Deformation of the cochlear bone then couples stronger to the basilar-membrane wave and produces a larger amplitude. 

The asymmetry between the two cochlear chambers, measured through the ratio $A_1/A_2$ of their cross-sectional areas, is another important factor in this mechanism as stated above (Fig.~\ref{WallStim}C). In a symmetric cochlea, deformation of the cochlear bone does not produce a deflection of the basilar membrane. In a real cochlea, however, the cross-sectional areas of both chambers differ. The evoked basilar-membrane vibration is the stronger the larger the asymmetry.

\newpage

\section*{Discussion}

Our results show that deformation of the cochlear bone can play a critical role for sound perception as well as for the propagation of otoacoustic emissions.  Deformation of the cochlear bone  can yield a fast wave, in addition to the much-studied slow basilar-membrane wave. Because the cochlea is asymmetric---the cross-sectional areas of both chambers differ---the two modes couple to each other. 

A force that acts on the basilar membrane, such as the one produced by the activity of hair cells, elicits not only a wave on the membrane, but  a wave on the cochlear bone as well. We have shown how distortion on the basilar membrane can accordingly  produce an otoacoustic emission that emerges from the inner ear through propagating from its generation site back to the stapes as cochlear-bone deformation. Because the wavelength of the cochear-bone mode is relatively long, on the order of a centimeter and hence comparable to the dimensions of the inner ear, the temporal delay of this emission is  small: the backward-propagating wave reaches the middle ear quickly. This mechanism can hence underlie the short-delay component of an otoacoustic emission.

Previously it has been suggested that the nonlinear distortion produced by  basilar-membrane vibration can launch a compressive fluid wave that propagates back to the stapes~\cite{Ren2004,hea-2007-228,He2008}. Our computations show that, when the cochlear bone is deformable, this wave does not only involve compression of the fluid but also deformation of the cochlear chambers. In fact, the latter effect dominates for the impedance associated to deformation of the cochlear bone is much less than that associated to compression of the fluid (equation~(\ref{Dispersion})). The wave accordingly has a significantly shorter wavelength than an ordinary compressive fluid wave.

The distortion in the cochlea also produces a reverse wave on the basilar membrane. Why has this component not been detected in recent laser-interferometric experiments? 

Our modeling reveals that a sizable portion of the backward-traveling wave on the cochlear bone becomes reflected at the middle ear and propagates forward, to the cochlear apex, both as a wave on the basilar membrane and as a cochlear-bone wave. We have quantified the magnitude of the reflected, forward-traveling basilar-membrane wave. For close primary frequencies as are typically used in experiments, the forward wave can have a significantly higher amplitude than the reverse basilar-membrane wave. Experiments will then only detect the forward-traveling wave. The stapes will accordingly vibrate before the basilar membrane, for the main component of basilar-membrane vibration arises from reflection at the stapes and hence occurs at a certain temporal delay. This delay has been measured in recent experiments~\cite{Ren2004}.

Our study shows that the backward-propagating basilar-membrane wave may dominate when the primary frequencies are sufficiently far apart. It will be interesting to see whether this reverse wave can indeed be experimentally measured, or whether its amplitude is too tiny for  distortion at far primary frequencies is small.

The one-dimensional model that we have employed cannot account for the drop in pressure near the peak of the basilar-membrane wave when deviating vertically  from the membrane. This pressure drop may alter the coupling to the cochlear-bone wave which may be interesting for future studies.

Stimulation of the cochlear bone---as elicited by bone-conduction headphones, for instance---can produce a basilar-membrane wave and accordingly yield a hearing sensation.  We have calculated the vibration of the basilar membrane and how it varies longitudinally. Our results show a basilar-membrane wave  that closely resembles the wave that emerges  from airborne sound. The amplitude is the stronger the larger the difference in cross-sectional areas of the two cochlear chambers. It also depends on the material properties of the cochlear bone. For realistic parameter values the amplitude of the membrane vibration corresponds to the experimentally-observed magnitude of bone conduction.

The increasing development and usage of bone-conduction headphones such as in as the Google glass device  and other commercial applications points to a need for a conceptual understanding of the underlying biophysics. We hope that the results we presented here help to clarify the mechanisms involved in bone conduction, and  to further advance its application.

\newpage

\section*{Methods}

\subsection*{Parabolic deflection of the basilar membrane}

We assume that each transverse segment of the  basilar membrane deflects parabolically. The membrane's width is $w_\text{bm}$, and we choose a transverse coordinate $y$ such that $y=-w_\text{bm}/2$ and $y=w_\text{bm}/2$ denote the points where the membrane segment is anchored in bone. The membrane velocity $V(y,t)$ is then
\begin{align}
V(y,t)=\frac{4V_\text{bm}}{w_\text{bm}^2}\left(y-\frac{w_\text{bm}}{2}\right)\left(y+\frac{w_\text{bm}}{2}\right),
\end{align}
in which $V_\text{bm}$ is the maximal basilar-membrane velocity (at its midpoint $y=0$).

The temporal changes $\partial_t a_1$ and $\partial_t a_2$ of the cochlear chambers' cross-sectional areas then follow as 
\begin{equation}
\partial_t a_1=-\partial_t a_2=\int_{-\frac{w_\text{bm}}{2}}^{\frac{w_\text{bm}}{2}} dy V(y,t),
\end{equation}
which yields equation~(\ref{eq:area_change}).

\subsection*{Linear-response coefficient C}

We consider a tube subject to radial pressure. The tube's wall is assumed to be incompressible and elastic such that the circumference of a cross-section of the tube remains constant under deformation. 

We assume that the cross-section of the tube is approximately elliptical, with a wall distance $r_0$ from the midpoint that depends on the central angle $\phi$ through $r_0(\phi)=R+w_0 \cos(2\phi)$ (Ref.\cite{TimoshenkoGere}, page 294-295). The variable $w_0$ hence measures the deviation of the cross-sectional shape from a circle, and the variable $R$ denotes the average wall distance. 

A change $p$ in the internal radial pressure  leads to a  deformation  $r(\phi)$ that we describe through a variable $w$: $r(\phi)=R+w \cos(2\phi)$. The magnitude of the change $\delta w=w-w_0$ is derived on page 295, Equations (7)-(18) in Reference~\cite{TimoshenkoGere}:
\begin{align}
\delta w=-\frac{w_0 p}{p_{cr}-p}    ,~~~\text{with}~~p_{cr}=\frac{E}{4(1-\nu^2)}\frac{h^3}{R^3}.
\end{align}
Here, $E$ denotes the Young's modulus of the cochlear bone, $\nu$ the Poisson ratio, and $h$ the thickness of the cochlear bone.

A small pressure change $p$ elicits an approximately proportional  change $\delta w$:
\begin{align}
\delta w&\approx \frac{4(1-\nu^2)}{E}\frac{R^3 w_0}{h^3}p.
\end{align}
A small change $\delta w$ in the variable $w$ leads, in turn, to a small area change. The area $A_\text{cs}$ of a cross-section can be computed from $r(\phi)$ as
\begin{equation}
A_\text{cs}=\frac{1}{2}\int_0^{2\pi}r^2d\phi.
\end{equation}
The area change $a$ follows, to first order in the change $\delta w$, as
\begin{equation}
a=\frac{\partial A_\text{cs}}{\partial w}\delta w=\frac{1}{2}\int_0^{2\pi}2r\frac{\partial r}{\partial w}\delta wd\phi=\pi w_0\delta w.
\end{equation}
The small pressure change $p$ hence induces an area change according to $a=Cp$, with the coefficient 
\begin{equation}
C=\frac{4\pi(1-\nu^2)}{E}\frac{R^3 w_0^2}{h^3}.
\end{equation}
The latter is the linear-response coefficient that we employ in Equations~(\ref{eq:area_deformation}).

\subsection*{Spatial impedance variation and WKB approximation}

The impedance of the basilar membrane varies systematically along the cochlea. The basilar-membrane wave accordingly changes its wavelength as it propagates from the base towards its resonant position. The change of the wavelength and the amplitude can be captured by the WKB approximation, which starts from the following ansatz for the pressures~\cite{lighthill-1981-106}:
\begin{align}
p_{1/2}=\tilde{p}_{1/2}(x)e^{i\omega t-\Phi_{1/2}(x)} + \text{c.c.}.
\end{align}
To fulfill the wave equation the amplitudes $\tilde{p}_{1/2}(x)$ and phases $\Phi_{1/2}(x)$ have to obey
\begin{align}
\partial_x^2\tilde{p}_{1/2}(x)+2i\partial_x \tilde{p}_{1/2}(x)\partial_x\Phi_{1/2}(x)-\tilde{p}_{1/2}(x)[\partial_x\Phi_{1/2}(x)]^2+i\tilde{p}_{1/2}(x)\partial_x^2\Phi_{1/2}(x)=-\tilde{p}_{1/2}(x)k(x)^2.
\end{align}
The real part $\partial_x^2\tilde{p}_{1/2}(x)+\tilde{p}_{1/2}(x)\{k(x)^2-[\partial_x\Phi_{1/2}(x)]^2\}=0$ implies that $\Phi_{1/2}(x)=\pm \int_{0}^x k(x')dx'$. The imaginary part,  $2 \partial_x\tilde{p}_{1/2}(x)\partial_x\Phi(x)+\tilde{p}_{1/2}(x)\partial_x^2\Phi(x)=0$, leads to $\tilde{p}_{1/2}(x)=\phi_i/\sqrt{k(x)}$. 

\subsection*{Green's functions}

Green's functions are pressures  that result from point-wise stimulation at $x_0$ along the cochlea at frequency $\omega$. Two types of Green's functions are important in our study. The first type, pressures $\tilde{p}_{1/2}^G(x|x_0,\omega)$, reflects stimulation of the basilar membrane. The second type,  pressures $\tilde{p}_{1/2}^W(x|x_0,\omega)$, arise from stimulating the cochlear bone.

We start with computing the Green's functions that result from a point force acting on the basilar membrane. Such a force appears in the 
 the boundary condition, equation~(\ref{eq:green}). 
We make the ansatz
\begin{align}
\tilde{p}_1^G (x|x_0,\omega)&=\int_{-\infty}^{\infty} dk G_1(k) e^{i\omega t-ik (x-x_0)},\cr
\tilde{p}_2^G (x|x_0,\omega)&=\int_{-\infty}^{\infty} dk G_2(k) e^{i\omega t-ik (x-x_0)},\label{GBasilarStim}
\end{align}
with the wave-vector dependent coefficients $G_1(k)$ and $G_2(k)$.
Using fluid-momentum equation~(\ref{eq:momentum}) with the continuity equation~(\ref{eq:continuity}) as well as equations~(\ref{eq:area_deformation}) and (\ref{eq:green}) we obtain two coupled ordinary differential equations,
\begin{align}
- i \omega \rho\left\{\frac{2 w_\text{bm}}{3 Z_\text{bm}(x)}\left[G_1(k) - G_2(k) - \frac{p_F}{2 \pi} \right] + 
      i \omega c G_1(k)\right\}  &= A_1 k^2G_1(k),\cr
			 i \omega\rho \left\{\frac{2 w_\text{bm}}{3 Z_\text{bm}(x)}\left[G_1(k) - G_2(k) - \frac{p_F}{2\pi}\right] -  i \omega c G_2(k)\right\} &= A_2 k^2 G_2(k).
\end{align}
The coefficients $G_1(k)$ and $G_2(k)$ follow as:
\begin{align}
G_1(k)&= \frac{i\omega p_F \rho (A_2 k^2 - c \omega^2 \rho) w_\text{bm}}{3 A_1 A_2 \pi Z_\text{bm}(x) L(k)},\cr
G_2(k)&= \frac{i\omega p_F \rho (-A_1 k^2 + c \omega^2 \rho) w_\text{bm}}{3 A_1 A_2 \pi Z_\text{bm}(x) L(k)}.
\end{align}
Here we have used the abbreviation  $L(k)=[2 i \omega \rho (F_1+F_2) w_\text{bm} + 3 F_1 F_2 Z_\text{bm}(x)]/[3 A_1 A_2 Z_\text{bm}(x)]$ with $F_{1/2}=A_{1/2} k^2 - c \omega^2 \rho$. $L(k)=0$  is the dispersion relation that we have derived earlier from the eigenvalues of the matrix $\mathcal{M}$, equation~(\ref{Dispersion}).

The Green's functions for bone stimulation can be derived analogously. Assume that both cochlear chambers, at a certain longitudinal location $x_0$, are sinusoidally compressed and expanded:
\begin{align}
\widetilde{a}_{1/2}&=\frac{2i}{3\omega}\cdot w_\text{bm}\cdot \tilde{V}_\text{bm}+C\cdot [\tilde{p}_{1/2}+p_F \cos(\omega t) \delta(x-x_0)].
\end{align}

We make the following ansatz for the Greens functions:
\begin{align}
\tilde{p}_1^W (x|x_0,\omega)&=\int_{-\infty}^{\infty} dk W_1(k) e^{i\omega t-ik (x-x_0)},\cr
\tilde{p}_2^W (x|x_0,\omega)&=\int_{-\infty}^{\infty} dk W_2(k) e^{i\omega t-ik (x-x_0)},\label{GWallStim}
\end{align}
which yields the amplitude equations
\begin{align}
- i \omega \rho\left[\frac{2 w_\text{bm}}{3 Z_\text{bm}} (p_1^W - p_2^W) + i c \omega \left(p_1 + \frac{p_F}{2 \pi}\right)\right] &= A_1k^2 p_1\cr
 i \omega \rho\left[\frac{2 w_\text{bm}}{3 Z_\text{bm}} (p_1^W - p_2^W) - c i \omega \left(p_2 + \frac{p_F}{2 \pi}\right)\right] &= A_2 k^2 p_2.
\end{align}
The solutions are 
\begin{align}
W_1(k)&=\frac{c \omega^2 p_F \rho [4 i \omega \rho w_\text{bm} + 3 A_2 k^2 Z_\text{bm}(x) - 
   3 c \omega^2 \rho Z_\text{bm}(x)]}{6 A_1 A_2 \pi Z_\text{bm}(x) L(k)},\cr
W_2(k)&=\frac{c \omega^2 p_F \rho [4 i \omega \rho w_\text{bm} + 3 A_1 k^2 Z_\text{bm}(x) - 
   3 c \omega^2 \rho Z_\text{bm}(x)]}{6 A_1 A_2 \pi Z_\text{bm}(x) L(k)}.
\end{align}
with $L(k)$ as  given above. In the symmetric case of equal chamber areas, $A_1=A_2$, we obtain $W_{1}(k)=W_{2}(k)$. No basilar-membrane displacement then arises for the pressures in both chambers are equal.

When attempting to compute the integral in the ansatz for the Green's functions, equations~(\ref{GBasilarStim}) and (\ref{GWallStim}), we encounter a problem: the integrand has a singularity at the wave vectors $k$ for which $L(k)=0$, that is, at those wave vectors that obey the dispersion relation. However, we can employ the residue theorem of complex analysis to compute the integrals. Indeed, for propagation apical of the generation site, that is at a location
 $x<x_0$, we can close the contour in the upper-half plane for the integrand there is exponentially suppressed. The integral then only involves a contribution from the poles in the upper-half plane. In the case of basilar-membrane stimulation, we obtain a contribution proportional to $[\partial_k W_{1/2}^{-1}(k)]^{-1}$. The pressures $p_{1/2}(-k_\text{bm},\omega,x_0)$ represent the pressures of the basilar membrane mode in the two chambers
\begin{align}
p_1(-k_\text{bm},\omega,x_0)&=2\pi i \frac{\sqrt{k_\text{bm}(x_0)}}{\sqrt{k_\text{bm}(x)}}\left(\frac{\partial}{\partial k} W_1(k)^{-1}\right)^{-1}\Big |_{k=-k_\text{bm}(x_0)}\cdot e^{i\int_{x}^{x_0} k_\text{bm}(x)dx+i\omega t}+c.c.,\cr
p_2(-k_\text{bm},\omega,x_0)&=2\pi i \frac{\sqrt{k_\text{bm}(x_0)}}{\sqrt{k_\text{bm}(x)}}\left(\frac{\partial}{\partial k} W_2(k)^{-1}\right)^{-1}\Big |_{k=-k_\text{bm}(x_0)}\cdot e^{i\int_{x}^{x_0} k_\text{bm}(x)dx+i\omega t}+c.c..
\end{align}
Analogous results can be obtained for the cochlear-bone wave with $k_{cb}(x)$.

In the opposite case, for a cochlear location basal to the generation site, $x>x_0$, the integration path can be closed in the lower-half plane. 

\subsection*{Middle ear pressure transformation}

The three ossicles of the middle ear---malleus, incus, and stapes---connect the ear drum to the oval window. Sound is accordingly transmitted from the ear canal to the cochlea, and can analogously be re-emitted from the  cochlea into the ear canal. How can these transfers be quantified?

Denote by $A_\text{a}$ and $A_\text{ow}$ the area of the tympanic membrane respectively the oval window, and by $l_\text{a}$ and $l_\text{w}$ the length of the mallus respectively the incus (Figure~1). The pressure in the ear canal is $p_3$, it acts on the tympanic membrane and produces an angular momentum $l_\text{a} A_\text{a} p_3$. The pressure $p_2$ in the upper cochlear chamber yields an angular momentum $l_\text{w} A_\text{ow} p_2$ which must match the first one:
\begin{equation}
l_\text{a} A_\text{a} p_3=l_\text{w} A_\text{ow} p_2.
\label{MiddleEar1}
\end{equation}

A second equation results from the fluid flows in the ear canal as well as in the upper cochlear chamber, $j_3$ and $j_2$, which must yield an equal angular deflection of the middle-ear bones:
\begin{equation}
\frac{j_3}{l_\text{a} A_\text{a}}=\frac{j_2}{l_\text{w} A_\text{ow}}.
\label{MiddleEar2}
\end{equation}
Finally, the pressure $p_1$ in the lower cochlear chamber creates a fluid flow $j_1$ at the round window that depends on its  impedance $Z_\text{rw}$:
\begin{equation}
p_1=Z_\text{rw} j_1.
\label{MiddleEar3}
\end{equation}
These three equations act as boundary conditions to the wave equations and allow to compute the extent to which a wave reaching the middle ear, either from the ear canal or from within the cochlea, is transmitted or reflected.

We first illustrate how this computation works by considering airborne sound traveling through the ear canal towards the tympanic membrane, with a wave vector $k_s=\omega\sqrt{\rho_\text{air}\kappa}$ in which $\rho_\text{air}$ and $\kappa$ are the air's density respectively compressibility.  Part of this wave will be reflected, such that the pressure in the ear canal is the sum of a forward- and a backward traveling sound wave:
\begin{equation}
p_3=\tilde{p}_{3,f}e^{i\omega t -i k_s x}+\tilde{p}_{3,b}e^{i\omega t +i k_s x}.
\end{equation}
Within the cochlea, forward-traveling waves on the basilar membrane (wave vector $k_\text{bm}$) as well as on the cochlear bone (wave vector $k_\text{cb}$) will be elicited:
\begin{equation}
p_{1/2}=\tilde{p}_{1/2,\text{bm}}e^{i\omega t -i k_\text{bm} x}+\tilde{p}_{1/2,\text{cb}}e^{i\omega t -i k_\text{cb} x}.
\end{equation}
The associated fluid flows at the middle ear can be obtained from Equations~(\ref{eq:momentum}) in which the cross-sectional areas are substituted by the corresponding membrane areas, namely the ones of the tympanic membrane, round and oval window. The boundary equations~(\ref{MiddleEar1}-\ref{MiddleEar3}) can then be solved for the amplitudes of the wave components:
\begin{align}
\tilde{p}_{3,b} &=\tilde{p}_{3,f}\frac{ [A_\text{ow} k l_\text{w}^2 \rho (H \omega \rho - A_\text{rw} K_2 Z_\text{rw}) + A_\text{a} l_\text{a}^2 \rho_\text{air} (-K_1 \omega \rho + A_\text{rw} H k_\text{bm} k_\text{cb} Z_\text{rw})]}{A_\text{ow} k l_\text{w}^2 \rho (H \omega \rho - A_\text{rw} K_2 Z_\text{rw}) + A_\text{a} l_\text{a}^2 \rho_\text{air} (K_1 \omega \rho - A_\text{rw} H k_\text{bm} k_\text{cb} Z_\text{rw})},\cr
\tilde{p}_{2,\text{bm}} &=\tilde{p}_{3,f}\frac{2 A_\text{a} A_1 k l_\text{a} l_\text{w} \rho (\omega \rho - A_\text{rw} k_\text{cb} Z_\text{rw})}{A_\text{ow} k l_\text{w}^2 \rho (H \omega \rho - A_\text{rw} K_2 Z_\text{rw}) + A_\text{a} l_\text{a}^2 \rho_\text{air} (K_1 \omega \rho - A_\text{rw} H k_\text{bm} k_\text{cb} Z_\text{rw})},\cr
\tilde{p}_{2,\text{cb}} &=\tilde{p}_{3,f}\frac{2 A_\text{a} A_2 k l_\text{a} l_\text{w}  \rho (\omega \rho - A_\text{rw} k_\text{bm} Z_\text{rw})}{A_\text{ow} k l_\text{w}^2 \rho (H \omega \rho - A_\text{rw} K_2 Z_\text{rw}) + A_\text{a} l_\text{a}^2 \rho_\text{air} (K_1 \omega \rho - A_\text{rw} H k_\text{bm} k_\text{cb} Z_\text{rw})}.
\end{align}
Here we have employed the following abbreviations: $K_1=A_1 k_\text{bm} + A_2 k_\text{cb}$, $K_2=A_2 k_\text{bm} + A_1 k_\text{cb}$, $H=A_1 + A_2$.

The middle ear matches impedances such that most of the energy of the sound wave is transmitted to the basilar-membrane wave. We employ this criterion to determine the impedance of the round window. Requiring that the incoming sound wave is not reflected at the middle ear but instead fully transmitted into the cochlea, we obtain the impedance of the round window as:
\begin{align}
Z_\text{rw}= \frac{\omega \rho (A_\text{ow} H k l_\text{w}^2 \rho - A_\text{a} K_1 l_\text{a}^2 \rho_\text{air})}{A_\text{rw} (A_\text{ow} k K_2 l_\text{w}^2 \rho - A_\text{a} H k_\text{bm} k_\text{cb} l_\text{a}^2 \rho_\text{air})}.
\end{align}

Next, we  consider how a distortion signal emerges from the cochlea through a cochlear-bone wave. To this end we compute how much of a backward cochlear-bone wave, as generated from distortion, is transmitted as a sound wave into the ear canal, and how much is reflected as forward-traveling wave in the cochlea (potentially both in the cochlear-bone and in the basilar-membrane mode).  We hence start from the following ansatz
\begin{align}
p_{1/2}&=\tilde{p}_{1/2,\text{cb},b} e^{i\omega t +i k_\text{cb} x}+\tilde{p}_{1/2,\text{cb},f} e^{i\omega t -i k_\text{cb} x}+ \tilde{p}_{1/2,\text{bm}} e^{i\omega t -i k_\text{bm} x},\cr
p_3&=\tilde{p}_{3} e^{i\omega t +i k_s x},
\end{align}
in which $\tilde{p}_{1/2,\text{cb},b}$ is the amplitude of the backward-propagating bone wave, $\tilde{p}_{1/2,\text{cb},f}$ the amplitude of the forward-traveling bone wave, $ \tilde{p}_{1/2,\text{bm}}$ the amplitude of the forward-propagating basilar-membrane wave, and $\tilde{p}_{3}$ the amplitude of the emitted sound wave. From Equations~(\ref{eq:momentum}) and (\ref{MiddleEar1}-\ref{MiddleEar3}) we compute those amplitudes as:
\begin{align}
\tilde{p}_3 &=\tilde{p}_{2,\text{cb},b}\frac{4 A_\text{a} A_\text{ow} H A_2 k k_\text{cb} l_\text{a}^2 l_\text{w}^2  \rho \rho_\text{air} (\omega \rho - 
    k_\text{bm} B)^2}{[A_\text{ow} k l_\text{w}^2 \rho (H \omega \rho - K_2 B) + 
  A_\text{a} l_\text{a}^2 \rho_\text{air} (K_1 \omega \rho -  H k_\text{bm} k_\text{cb} B)]^2},\\
\tilde{p}_{2,\text{bm}} &=	\tilde{p}_{2,\text{cb},b}\frac{4 A_\text{a} A_1 A_2 k k_\text{cb} l_\text{a} l_\text{w} \rho^2 (\omega \rho - 
    k_\text{bm} B) (A_\text{a} l_\text{a}^2 \omega \rho_\text{air} + 
    A_\text{ow} k l_\text{w}^2 B)}{[A_\text{ow} k l_\text{w}^2 \rho (H \omega \rho - K_2 B) + 
  A_\text{a} l_\text{a}^2 \rho_\text{air} (K_1 \omega \rho -  H k_\text{bm} k_\text{cb} B)]^2},\cr
	\tilde{p}_{2\text{cb},f} &= -\tilde{p}_{2,\text{cb},b}\frac{2 A_\text{a} A_2 k l_\text{a} l_\text{w}  \rho (\omega \rho - k_\text{bm} B) [A_\text{a} l_\text{a}^2 \rho_\text{air} (K_3 \omega \rho +  H k_\text{bm} k_\text{cb} B) + A_\text{ow} k l_\text{w}^2 \rho (H \omega \rho + 
        K_4 B)]}{[A_\text{ow} k l_\text{w}^2 \rho (H \omega \rho - 
       K_2 B) + A_\text{a} l_\text{a}^2 \rho_\text{air} (K_1 \omega \rho - H k_\text{bm} k_\text{cb} B)]^2}.\nonumber
\end{align}
In addition to the abbreviations introduced above, we have used the following: $B=A_\text{rw} Z_\text{rw}$,  $K_3=A_1 k_\text{bm} - A_2 k_\text{cb}$, $K_4=A_1 k_\text{cb} - A_2 k_\text{bm}$, $K_s=k_\text{bm} + k_\text{cb}$, and $K_d=k_\text{bm} - k_\text{cb}$.

\newpage



\newpage

\section*{Acknowledgments}
We would like to thank A.~J.~Hudspeth and L.~Abbott for helpful discussions and the members Center for Theoretical Neuroscience at Columbia University for hospitality (T.~T.). This work has been supported by the the Max Planck Society and the Volkswagen Foundation through a Computational Sciences fellowship (to T.~T.) and by a Career Award at the
Scientific Interface from the Burroughs Wellcome Fund (to T.~R.).

\section*{Author contributions}

T.~T. and T.~R. planned the research, analyzed the data, and wrote the article. The analytical and numerical computations were performed by T.~T..

\section*{Competing financial interests}

The authors have no competing financial interests.

\newpage

{\bf \Large Figures}

\vspace*{5cm}

\newpage

\begin{figure}[h]
\includegraphics[width=0.5\columnwidth]{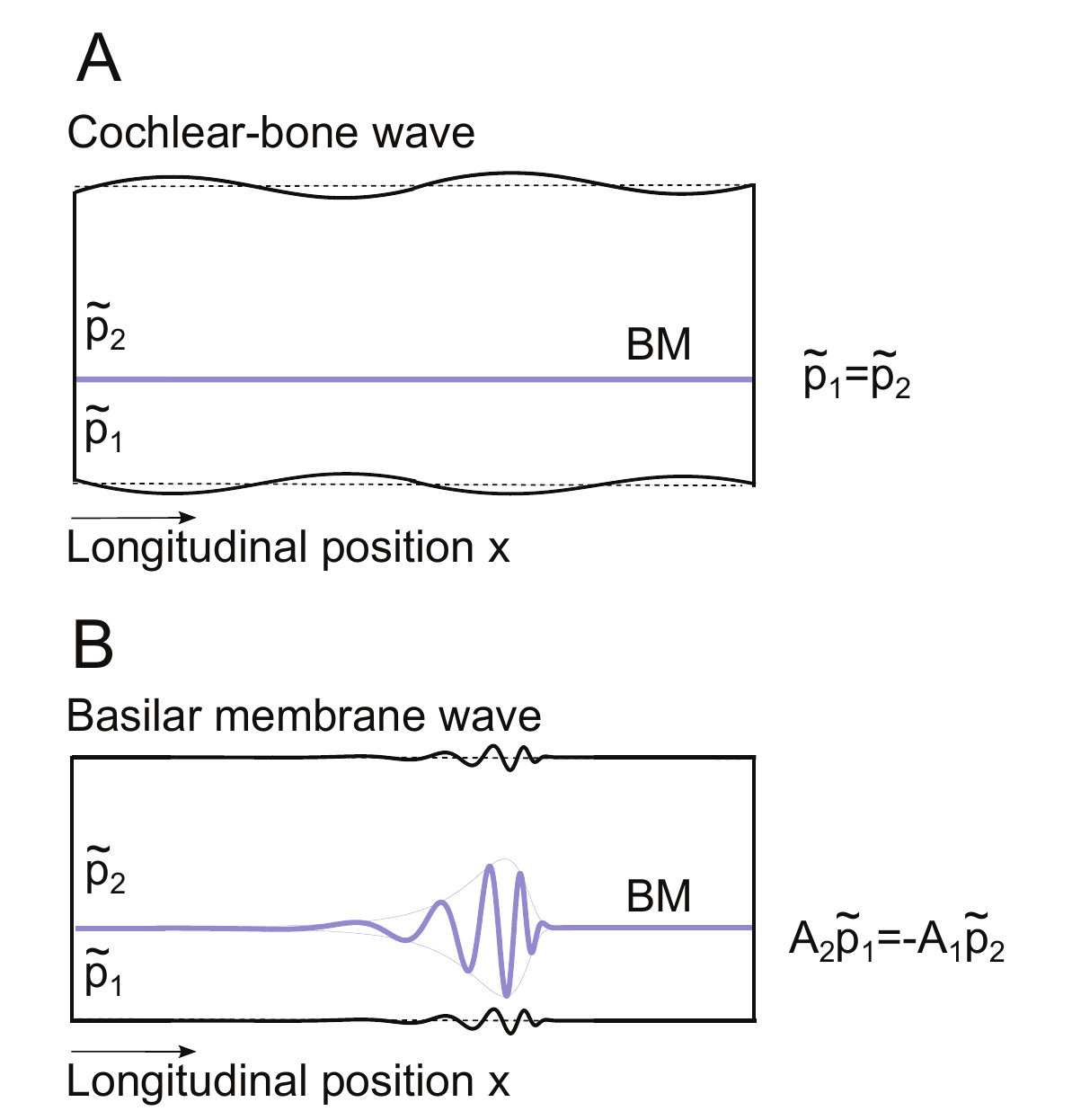}
\caption{{\bf Basilar-membrane and cochlear-bone wave} Two independent propagation modes, the basilar-membrane wave and the cochlear-bone wave, exist in a cochlea with a deformable bone. ({\bf A}) Deformation of the cochlear bone propagates longitudinally as a wave that elicits, at a given location, identical pressure changes in both chambers and hence no vibration of the basilar membrane. ({\bf B}) The basilar-membrane wave is evoked by a pressure difference across it. Because of the pressure changes in each chamber, this wave is accompanied by deformation of the cochlear bone.
\label{Modes}}
\end{figure}

\vspace*{3cm}

\begin{figure}[h]
\centering
\includegraphics[width=0.7\columnwidth]{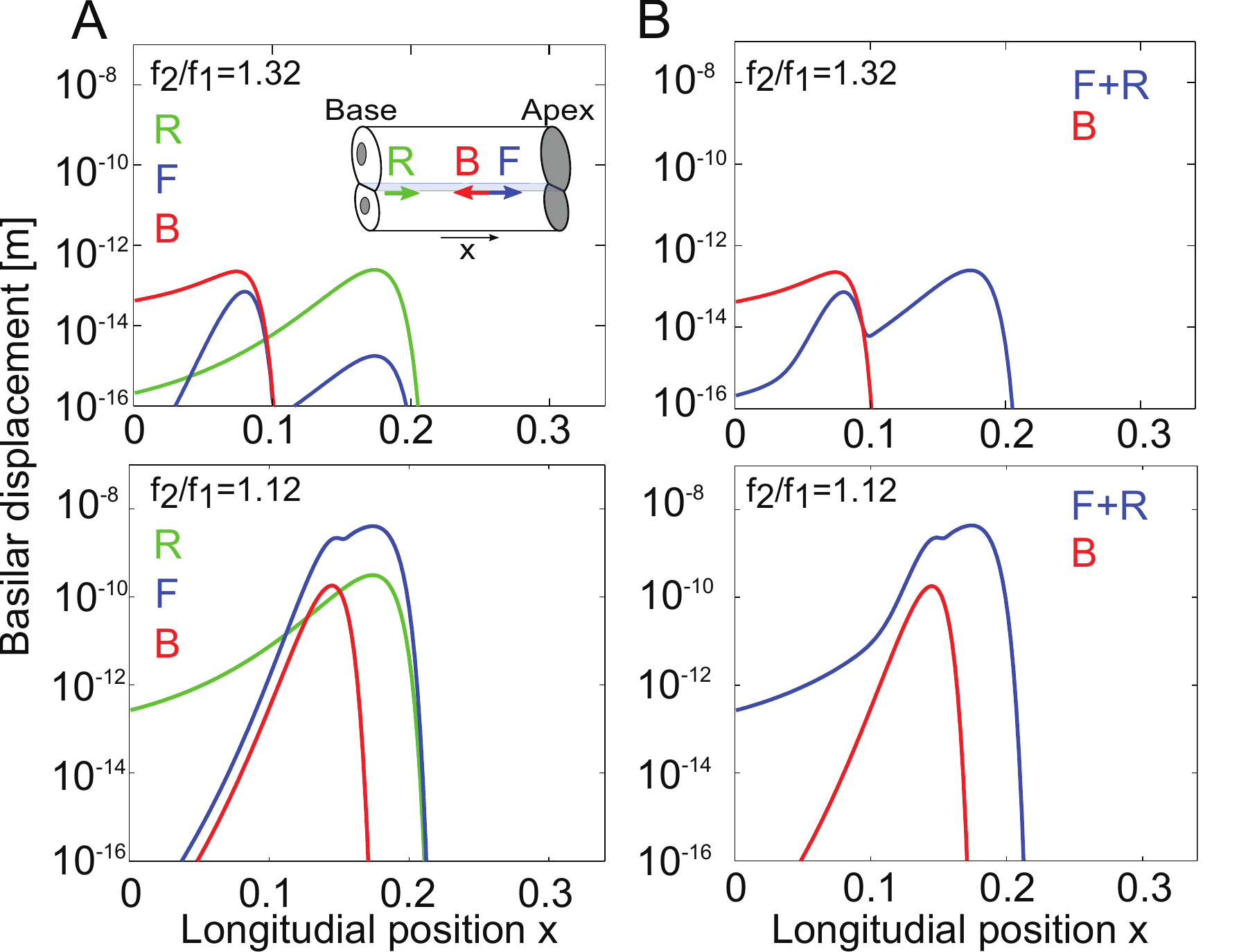}
\caption{{\bf Distortion-product waves traveling towards and away from the base} ({\bf A}) Amplitude of basilar-membrane waves  at distortion frequency $f=2 f_1-f_2=7$ kHz. One component travels backward from the generation site toward the base (`B', red)  whereas another propagates forward to the apex (`F', blue). A third wave emerges from the reflection of the backward-propagating cochlear-bone wave and travels forward on the membrane (`R', green). The three contributions are illustrated in the inset. The upper and lower panel show the amplitudes for two different ratios of the primary frequencies. ({\bf B}) The reflected and the forward-traveling basilar-membrane wave combine to a net forward-traveling wave (`F+R', blue). Depending on the ratio of the primary frequencies as well as the cochlear location, this wave can overwhelm the backward-propagating one (`B', red).
\label{ForwardBackward}}
\end{figure}

\newpage

\vspace*{3cm}

\begin{figure}[h]
\includegraphics[width=\columnwidth]{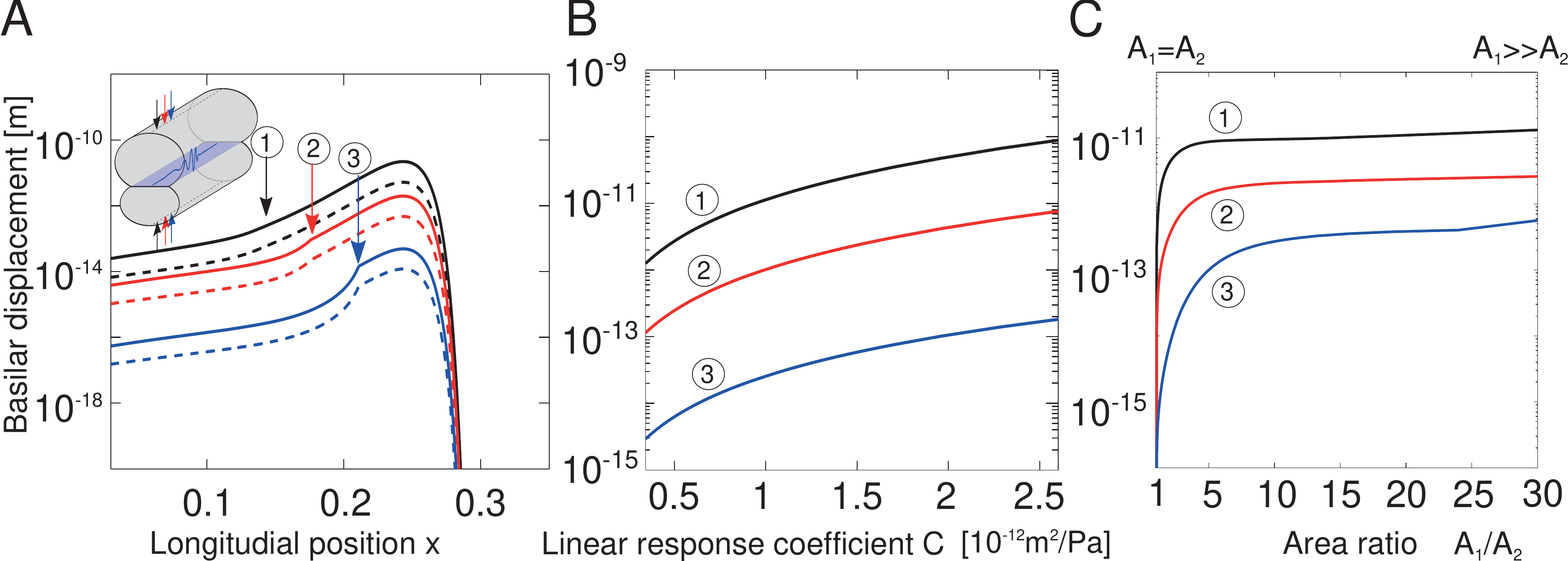}
\caption{{\bf Basilar-membrane excitation through cochlear-bone compression} ({\bf A}) Deformation of the cochlear chambers at different location $x$, illustrated by the numbered vertical arrows, can deflect the basilar membrane. The inset schematically shows the compressive stimulation of the cochlear bone; cochlear bone linear response $C=6.86\cdot 10^{-13}$~m$^2$/Pa. ({\bf B}) The maximal basilar-membrane deflection depends on the material properties of the cochlear bone. It is the stronger the larger the linear response coefficient of the bone, for that implies a smaller bone impedance  which then is closer to that of the basilar membrane. ({\bf C}) The maximal basilar-membrane deflection depends  on the ratio $A_1/A_2$ of the cross-sectional areas $A_1$ and $A_2$ of the two chambers. The membrane displacement vanishes in a symmetric cochlea ($A_1/A_2=1$) and grows with increasing asymmetry ($A_1\ll A_2$).}
\label{WallStim}
\end{figure}

\clearpage
\newpage

{\bf \Large Table}
\vspace*{1cm}

\begin{table}[h]
\centering
\caption{Summary of model parameters}
\hspace*{-1.5cm}
\begin{tabular}{llll}
Quantity&Description& Value& Citation\\
\hline
$A_\text{ow}$&Area of the oval window& 2.3 mm$^2$&\cite{Kringlebotn2000,Pickles}\\ 
$A_\text{rw}$&Area of the round window&$2 A_\text{ow}$&\cite{Kringlebotn2000}\\
$A_\text{a}$&Area of the tympanic membrane& $35 A_\text{ow}$&\cite{Pickles}\\
$A_1/A_2$& Area ratio between the two cochlear chambers&$0.42$& \cite{CochleaAnatomy}\\
$A_2+A_1=$const& Total area of the two cochlear chambers&$1200$~mm$^3$& \cite{Pickles}\\
$l_\text{w}$&Incus length&$4 \cdot 10^{-3}$ m&\cite{Pickles} \\
$l_\text{a}$&Mallus length&1.15 $l_\text{w}$ &\cite{Pickles}\\
$\rho$&Cochlear fluid density&$1000$ kg/m$^3$& \cite{Pickles} \\
$\rho_\text{air}$&Air density&$1.2$ kg/mm$^3$& \cite{Pickles}\\ 
$h$& Thickness of the cochlear bone&$0.01\cdot 10^{-3}$ m &\cite{Pickles}\\
$\nu$& Poisson ratio of cochlear bone&$0.3$&\cite{Pickles,TimoshenkoGere}\\
$E$&Young's modulus of the cochlear bone &$27.8\cdot 10^9$ kg/m/s$^2$& \cite{Pickles,YoungModuli,TimoshenkoGere} \\
$R$& Average  radius of a cochlear chamber&$6\cdot 10^{-4}$ m& \cite{Pickles} \\
$w_0$& Elliptical deformation of a cochlear chamber&$1\cdot 10^{-4}$ m& \cite{Pickles} \\
A& Strength of the nonlinear membrane response&$5\cdot10^{23}$& \\
$w_\text{bm}(x)$& Width of the basilar membrane& $10^{-6}(100 + 400 x)$ m& \cite{Pickles,reichenbach-2012-1}\\
$A_\text{bm}(x)$& Area of a basilar-membrane segment&$w_\text{bm}(x)\cdot 8\mu$ m& \cite{lighthill-1981-106,reichenbach-2012-1}\\
$K(x)$& Stiffness of the basilar membrane & $f_0(x)/f_0(0)$ N/m& \cite{lighthill-1981-106,reichenbach-2012-1}\\
$f_0(x)$&Resonant frequency of the basilar membrane&$30\cdot10^3\cdot e^{-\log(30\cdot 10^3/50)x}$ Hz& \cite{lighthill-1981-106,reichenbach-2012-1}\\
$m(x)$&Mass of the basilar membrane &$K(x)/(2\pi f_0(x))^2$ & \cite{lighthill-1981-106,reichenbach-2012-1}\\
$\mu(x)$&Drag coefficient of the basilar membrane&$w_\text{bm}(x)\cdot 0.015$ Ns/m$^2$& \cite{lighthill-1981-106,reichenbach-2012-1}\\
$Z_\text{bm}(x)$&Basilar membrane impedance & $\frac{1}{A_\text{bm}(x)}\left[\frac{-i K(x)}{\omega}+\mu(x)+i\omega m(x)\right]$& \cite{lighthill-1981-106,reichenbach-2012-1}\\
\end{tabular}
\label{Parameters}
\end{table}

\end{document}